\documentclass[11pt,a4paper]{scrartcl}

\usepackage{ILD}

\usepackage[symbol]{footmisc}
\usepackage{feynmf}
\usepackage{textpos}
\usepackage{lineno}
\usepackage{wrapfig}

\title{Status of $e^+e^-$ Higgs Factory Projects}

\date{\today}

\addauthor{Jenny List}{\institute{1}}

\addinstitute{1}{Deutsches Elektronen-Synchrotron DESY, Notkestr. 85, 22607 Hamburg, Germany}

\abstract{An electron-positron collider designed for precision studies of the Higgs boson, a so-called Higgs factory is the highest-priority next collider of the particle physics community. This contribution summarises the key physics goals of such a Higgs factory and reviews the status of the various proposed realisations from mature concepts to very recent ideas. The commonalities and special advantages of circular and linear approaches will be discussed, respectively, before highlighting some recent developments regarding the key technologies, the operation scenarios and sustainability aspects for future colliders. }

\titlecomment{Talk presented at the European Physical Society Conference on High Energy Physics (EPS-HEP2023), 21-25 August 2023, Universit\"at Hamburg, Hamburg, Germany.}

\graphicspath{ . }

\begin{document}

\titlepage

\section{Higgs Factory Projects}
\begin{figure}[htb]
\begin{center}
\includegraphics[width=0.5\textwidth]{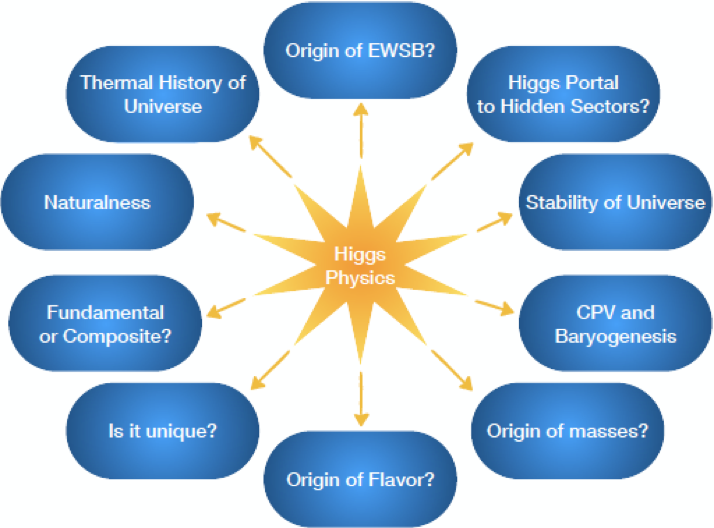}
\caption{The Higgs boson is at the center of many fundamental questions on our universe and its evolution. From~\cite{Narain:2022qud}.}
\label{fig:higgsstar}
\end{center}
\end{figure}

In the last update of the European Strategy for Particle Physics, an $e^+e^-$ collider serving as a Higgs factory was identified as the highest priority next collider~\cite{EPPSU2020}. Last year, this was re-emphasised in the final report of the Snowmass Community Study in the US, recognising that the Higgs boson is intimately connected to nearly all our fundamental questions about the origin and development of the universe, as illustrated in Fig.~\ref{fig:higgsstar}, and that thus a significantly more precise characterisation of its properties and its potential is an essential ingredient to addressing these questions~\cite{Butler:2023glv}. 

This contribution summarises the status of the main $e^+e^-$ Higgs factory projects, FCCee, CEPC, CLIC and ILC, as well as of more recent proposals like C3, following a series of recent dedicated workshops on these projects~\cite{LCWS2023, FCC2023, CEPC2023, CCC2023}.  The most mature among these proposals, the International Linear Collider, ILC~\cite{ILCInternationalDevelopmentTeam:2022izu, Behnke:2013xla, ILC:2013jhg, Adolphsen:2013jya, Adolphsen:2013kya, Behnke:2013lya}, is based on superconducting RF as used successfully in e.g.\ the EU-XFEL and LCLS-II. In its Higgs Factory stage at a center-of-mass energy ($\sqrt{s}$) of $250$\,GeV, it would be about $20$\,km long and could be operated also at lower energies down to the $Z$ pole, while being upgradable to $\sqrt{s}=1$\,TeV. Since several years, it is under political consideration for construction in Japan, where most recently the ILC Technology Network was launched to address the last remaining R\&D questions. The Compact Linear Collider, CLIC~\cite{Linssen:2012hp, CLICdp:2018cto}, could reach $\sqrt{s}=380$\,GeV with a footprint of only $11$\,km thanks to the drive-beam technology pioneered at CERN, and could eventually be upgraded to $3$\,TeV.

CERN's first priority in terms of future projects, however, is the Future Circular Collider, FCC~\cite{FCC:2018evy}, a $90$\,km circular facility which could host an $e^+e^-$ collider running at the $Z$ pole, the $WW$ threshold, the $ZH$ threshold and eventually the $t\bar{t}$ threshold, before being replaced by a proton-proton collider. Currently, the feasibility of such a project is being studied, with a final report expected by the end of 2025. A similar project, the Circular Electron Positron Collider, CEPC~\cite{CEPCStudyGroup:2018rmc, CEPCStudyGroup:2018ghi} is being developed in China. CEPC is in the final phases of publishing a Technical Design Report, including a costing reviewed by international experts, and aims for an approval in the next 5-year plan in 2025. If indeed a decision to construct CEPC or ILC is taken in the next couple of years, these machines could start data-taking in the late 2030ies, with overlap to HL-LHC operation. Due to the latter, any new project at CERN could not start data-taking before the end of 2040ies.

\begin{figure}[htb]
\begin{center}
\includegraphics[width=0.75\textwidth]{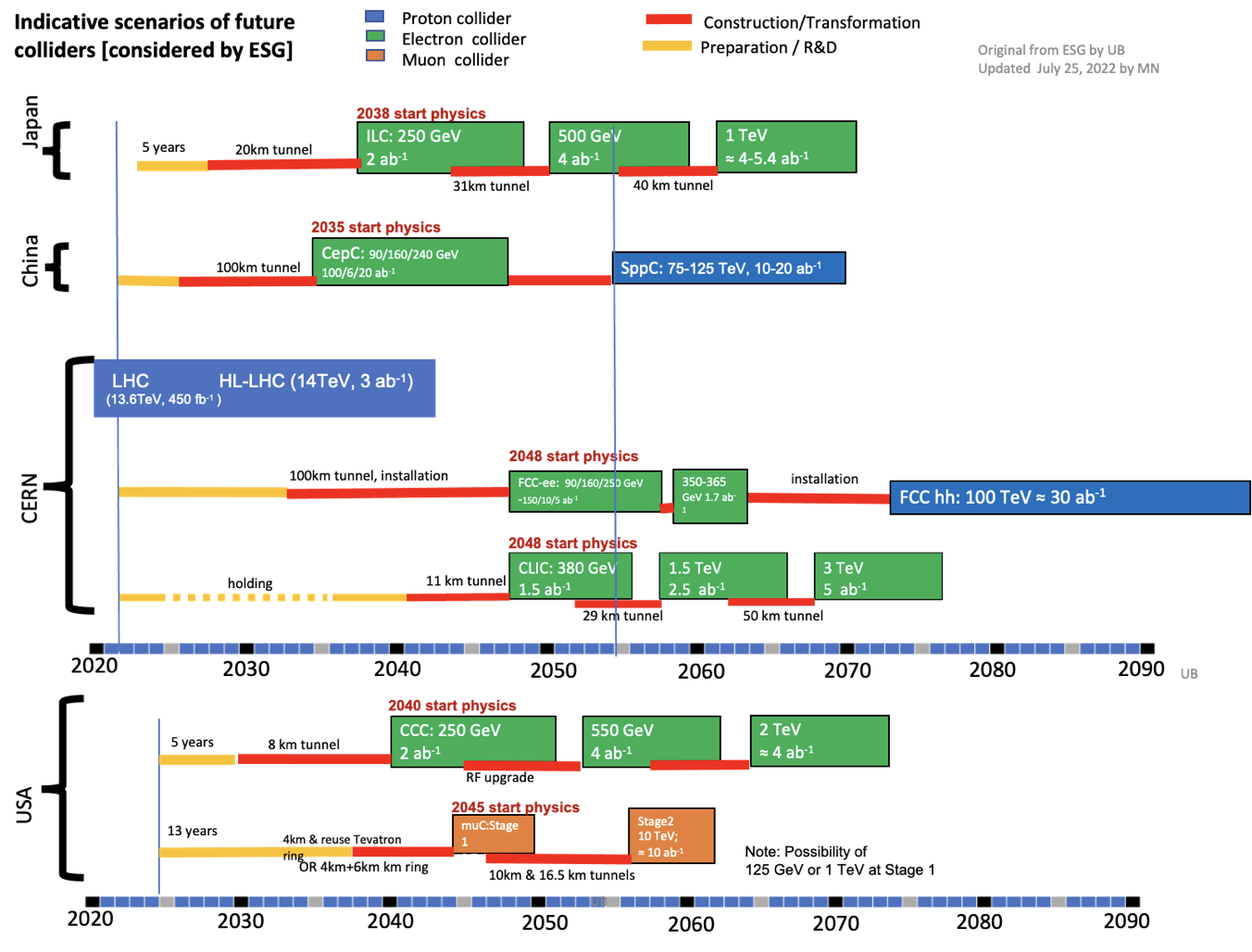}
\caption{Envisioned timeslines of various future collider projects as discussed in the Snowmass process. For most projects, they are technically driven and would allow a start of data-taking in the late 2030ies -- if a decision for construction is made soon. The timeline of the CERN-based projects is coupled to the completion of the HL-LHC programme, shifting the possible start of data-taking into the late 2040ies.}
\label{fig:timelines}
\end{center}
\end{figure}
More recently, a linear collider based on cooled copper cavities, the Cool Copper Collider, C3~\cite{Dasu:2022nux}, has been proposed in the US. Thanks to much higher gradients than possible in SCRF, it requires only a footprint of about $4$\,km to reach $\sqrt{s}=250$\,GeV, and even a $550$\,GeV version would still fit into the FNAL site. Assuming a significant investment in the required R\&D over the next five years, C3 could start operation in the early 2040ies. Figure~\ref{fig:timelines} summarizes the current view on potential timelines of various future colliders.

There is even a newer kid on the block: only in spring this year, a Hybrid Asymmetric Linear Higgs Factory, HALHF ~\cite{Foster:2023bmq}, has been proposed based on colliding a high-energy, plasma-wakefield accelerated electron beam with a lower energy conventional positron beam. This idea motivates a significant plasma acceleration R\&D program targeting collider applications, in particular to demonstrate the concatenation of several plasma cells.

While all these proposals differ in the accelerating technology and the site-specific design, they fall into two basic categories: circular colliders and linear colliders, each with their own advantages. Circular colliders offer several interaction regions, very clean experimental conditions and very high luminosity and power-efficiency at their lower energy stages. Linear colliders have a smaller footprint, offer polarised beams and give high luminosity and power-efficiency at high energies. Both also come with visions for the long-term future: circular facilities could host a proton-proton collider, while linear facilities could be upgraded in energy -- either by extending the tunnel, or by replacing the accelerating structures with advanced accelerator technologies.

\section{Physics Considerations}

\begin{figure}[htb]
\begin{center}
\includegraphics[width=0.95\textwidth, clip, trim = 0 255 0 0]{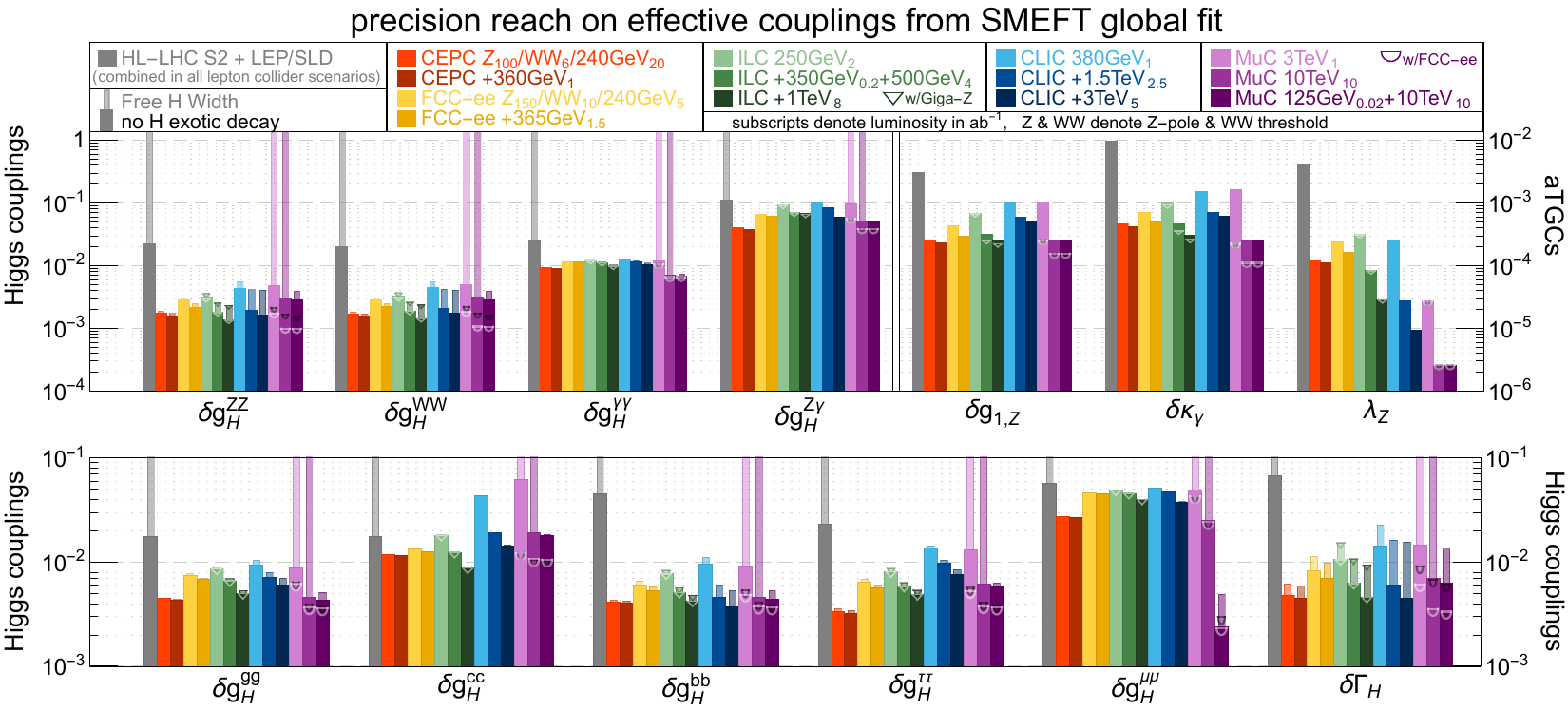}
\caption{Constraints on Higgs and triple gauge couplings from SMEFT fit to Higgs and electroweak sector for HL-LHC and various proposed lepton colliders~\cite{deBlas:2022ofj}. Compared to HL-LHC and the MuonCollider, all $e^+e^-$ projects deliver a comparable performance.}
\label{fig:rainbow}
\end{center}
\end{figure}
Despite these different features, all the proposed projects perform surprisingly similarly with respect to the core Higgs factory programme, as shown in the most recent SMEFT evaluation~\cite{deBlas:2022ofj} performed for the Snowmass Community Study, and displayed in Fig.~\ref{fig:rainbow}: the Higgs boson's couplings to the $Z$ and $W$, to the gluons as well as to $b$ quarks and $\tau$ leptons will be constrained at the level of a few permille, and the couplings to $c$ quarks and photons at the level of one percent.

\begin{figure}[tb]
\begin{center}
\includegraphics[width=0.5\textwidth]{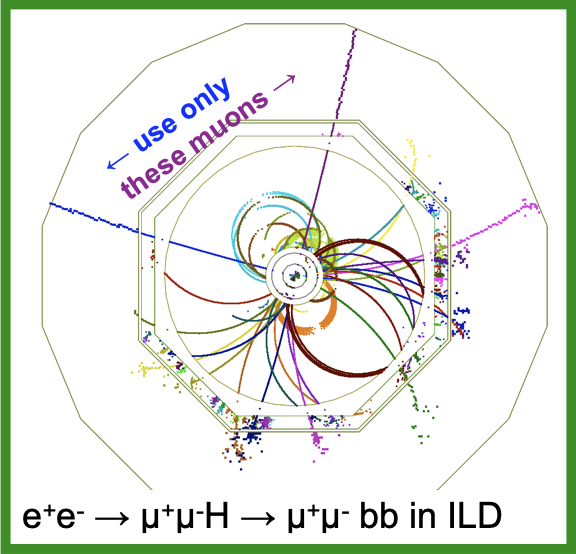}
\caption{Event display of $e^+e^- \to ZH \to \mu^+\mu^- b \bar{b}$ in the ILD detector~\cite{bib:ILDevents}}
\label{fig:ZHevt}
\end{center}
\end{figure}

The most striking gain over the HL-LHC capabilities, however, is that the absolute couplings become accessible without any assumptions on the total width or the absence of exotic decay modes. Thanks to the known initial-state four-momentum, $e^+e^-$ colliders offer the unique possibility to detect $e^+e^- \to ZH$ events only via the recoil against the $Z$ boson, independently of the decay modes of the Higgs boson, as illustrated by a simulated event display in Fig.~\ref{fig:ZHevt}. This model-independent determination of the absolute couplings represents a qualitative leap in terms of knowledge and in our ability to test extensions of the Standard Model.

Beyond the minimal Higgs programme, however, circular and linear colliders do place emphasis on different areas of physics. The very high luminosity of circular machines at the $Z$ pole offers a rich programme of physics at the intensity frontier, ranging from electroweak and QCD precision measurements to $b$- and $\tau$-physics and searches for very light exotic particles, but also creates challenges for the detectors. Linear colliders, on the other hand, offer a unique programme at higher centre-of-mass energies, comprising di-Higgs production, direct searches for new particles and a full precision programme for the top-quark, including the full CP structure of its electroweak and Yukawa couplings. A recent SMEFT fit of the top sector based on projections for the HL-LHC and various $e^+e^-$ projects clearly showed that $e^+e^-$ collisions at energies of $500$\,GeV and beyond with polarised beams are required in order to lift degeneracies between operators~\cite{Durieux:2022cvf}.

\section{Recent Developments}
During the Snowmass process, a task force analysed the technical readiness, costs and timelines of the various proposed future projects~\cite{Roser:2022sht}. The radio-frequency systems and the positron source were identified as critical components limiting the technological readiness level of the main Higgs factory contenders --- and at the same time, a lot of progress is being achieved in these areas.

Concerning SCRF cavities, which are not only needed for the ILC main linac, but also for the booster and collider rings of circular colliders as well as the damping rings of linear colliders, very interesting progress has been shown recently in the new PAPS facility~\cite{bib:paps} at IHEP, reaching very high quality factors of about $5\cdot 10^{10}$ at a gradient of $31$\,MV/m with a simplified surface treatment receipe~\cite{bib:highQ}. The performance surpasses the CEPC specs and promises lower power consumption and reduced production costs at the same time. An other important means to reduce the power consumption is the development of more efficient klystrons, where both CERN and IHEP are reaching efficiencies beyond $80\%$ based on new design methods~\cite{Zhou:2023poz, Marchesin:2018ubn}. The power estimates for CLIC, FCCee and CEPC are based on these developments, while the power estimates for ILC still assume an efficiency of 65\% based on commercially available klystrons. 

SCRF cavities and the positron source are also two important topics of the ILC Technology Network~\cite{bib:IDT-EB-2023-001, bib:IDT-EB-2023-002}, which recently was launched by an agreement between CERN and KEK. Within the ITN, CERN will act as a European hub to facilitate money transfer to other labs and universities. As a first project, design and prototyping of a new capture device, a pulsed solenoid, for the positron source is being pursued, since simulations showed recently the potential to increase the positron yield substantially with such a device~\cite{bib:positron}. 

\begin{figure}[tb]
\begin{center}
\includegraphics[width=0.75\textwidth]{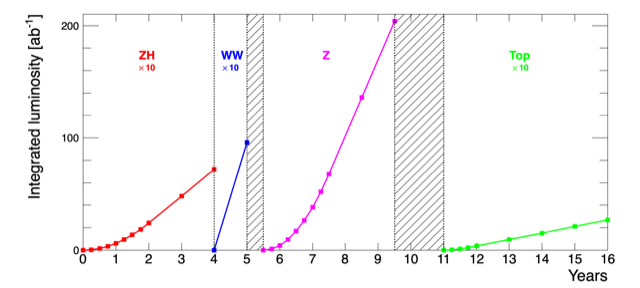}
\caption{Alternative running scenario for FCCee starting with a Higgs run.}
\label{fig:fccee_run}
\end{center}
\end{figure}

There are also new ideas on how to operate these machines: FCCee is now considering the possibility of starting operation with the Higgs physics run, as illustrated in Fig.~\ref{fig:fccee_run} and later installing the dedicated cavities for the high-luminosity run at the $Z$ pole, which adds important flexibility to the running programme. CEPC is investigating the possibility of providing a longitudinally polarised electron beam to the experiments~\cite{Assmann:2023ijx}. Simulations show that when using a polarised source with $|P| > 85\%$, up to $70\%$ transverse polarisation could be maintained through the booster ring and the top-up injection into the collider ring. The next steps will be to integrate spin rotators and polarimeters into the lattice. 

When looking at the running scenarios for linear colliders, a peculiar situation can be found: While ILC historically picked $500$\,GeV, it has been known ever since the discovery of the Higgs boson at a mass of $125$\,GeV that this is slightly too low for an optimal study of $t\bar{t}H$ production~\cite{Barklow:2015tja}. Therefore, C3 chose $550$\,GeV as the baseline, which immediately improves the projections for the achievable precision on the top Yukawa coupling by more than a factor of two. The CLIC running programme doesn't consider data-taking in the range of $500$ to $600$\,GeV at all, and instead proposes to jump directly from $380$\,GeV to $1.5$\,TeV. This seems to indicate an emerging need to re-discuss the optimal physics choice of energy stages beyond the Higgs run.

\section{Sustainability}
\begin{figure}[htb]
\begin{center}
\includegraphics[width=0.6\textwidth]{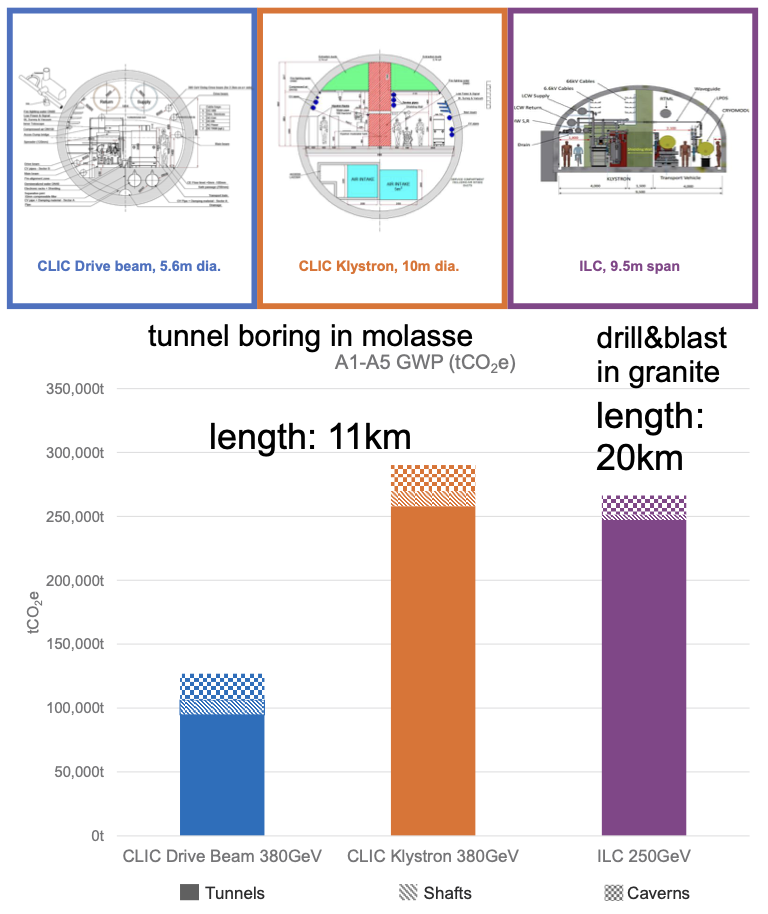}
\caption{Comparison of the tunnel shapes, length, tunneling techniques for ILC250 and the driven-beam and klystron-based options for CLIC380.}
\label{fig:arup}
\end{center}
\end{figure}

Last but not least, the particle physics community has realised that any future project needs to be designed and chosen taking into account its overall resource consumption and ecological footprint. Based large share of the global warming potential (GWP) of future colliders is due to the construction of the tunnel, in particular the production of concrete and steel. As future circular colliders have a rather large circumference, the tunnel construction seems to dominate the GWP, while for linear colliders, which have a shorter footprint, the operation dominates~\cite{Breidenbach:2023dac}. On the positive side, a full life-cycle assessment according to ISO standards recently performed for CLIC and ILC~\cite{bib:arup} showed the potential to reduce the GWP in the construction by up to $40\%$ by using low-CO2 materials and by optimising the thickness of the tunnel walls. Also the tunnel shape plays an important role, as illustrated in Fig.~\ref{fig:arup}: While the 5.6\,m diameter round tunnel for the CLIC drive-beam baseline has a very similar CO2 footprint per kilometer as the kamaboko-shaped ILC tunnel with a span of 9.5\,m, the 10\,m diameter tunnel foreseen for the klystron option of CLIC produces more than twice as much CO2 per kilometer.

Realising the importance of sustainability considerations for future projects, the European Laboratory Directors Group recently established a task force to develop coherent criteria and guidelines for an assessment of new infrastructures. This group is expected to deliver a final set of standards by summer 2024, to be used by all proposals for the next update of the European Strategy for Particle Physics~\cite{bib:ldg_sus}.

\section{Conclusions}
There is a strong agreement in the world-wide particle physics community precision studies of the Higgs boson are crucial to answer our burning questions about the universe and its evolution. An electron-positron collider operating as a Higgs factory as therefore been identified as the highest-priority next collider. Several concrete projects have been proposed, all being able to deliver the core Higgs factory program, but being complementary in their further ambitions. Studies of the expected physics performance are essential to understand better the implications of the choices to be made. An active R\&D program is addressing the last remaining technological challenges, as well as working towards reducing the ecological footprint of these machines.

The approval of any new large-scale collider, however, requires a strong HEP community behind it, and an excellent communication of the scientific case to experts from other fields, funding agencies and the general public, disseminated by a growing community of supporters. 
Thus: get engaged and make it happen!

\end{document}